# Does the use of open, non-anonymous peer review in scholarly publishing introduce bias? Evidence from the F1000 post-publication open peer review publishing model

Mike Thelwall, Statistical Cybermetrics Research Group, University of Wolverhampton, Wulfruna Street, Wolverhampton WV1 1LY, UK. ORCID:0000-0001-6065-205X
Verena Weigert, Jisc, 15 Fetter Lane, London EC4A 1BW. ORCID: 0000-0003-4887-7867
Liz Allen, F1000, 34-42, Cleveland Street, London, W1T 4LB.
ORCID: 0000-0002-9298-3168
Zena Nyakoojo, F1000Research, London W1T 4LB, UK. ORCID: 0000-0001-6812-4120
Eleanor-Rose Papas, Editorial Department, F1000Research, London W1T 4LB, UK. ORCID: 0000-0003-4293-4488

**Background**: Peer review remains an important feature of scholarly publishing to assure scrutiny and trust in research findings. As part of moves towards open knowledge practices, making peer review more open is increasingly cited as a way to deliver even fuller scrutiny and transparency of assessments around research – and there are many flavours of open peer review now in use across scholarly publishing. Open peer review, particularly where reviews are *fully attributable and non-anonymised*, has however been subject to criticism for the bias that may accompany its use. This study examines whether there is any evidence of bias in two areas of common critique of open, non-anonymous peer review – and used in the post-publication, peer review system operated by the open-access scholarly publishing platform F1000Research. **First**, is there evidence of bias where a reviewer based in a specific country assesses the work of an author also based in the same country? **Second**, are reviewers influenced by being able to see the comments and know the origins of previous reviewer?
**Methods**: Scrutinising the open peer review comments published on F1000Research, we assess the extent of two frequently cited potential influences on reviewers that may be the result of the transparency offered by a fully attributable, open peer review publishing model: the national affiliations of authors and reviewers, and the ability of reviewers to view previously-published reviewer reports before submitting their own. The effects of these potential influences were investigated for all first versions of articles published between July 13, 2012, and July 8, 2019, to F1000Research.
**Results**: First, in 16 out of the 20 countries with the most articles, there was a tendency for reviewers based in the same country to give a more positive review; the difference was statistically significant in one. Only 3 countries had the reverse tendency. Second, there is no evidence of a conformity bias. When reviewers mentioned a previous review in their peer review report, they were not more likely to give the same overall judgement. Although reviewers who had longer to potentially read a previously published reviewer reports were slightly less likely to agree with previous reviewer judgements, this could be due to these articles being difficult to judge rather than deliberate non-conformity.
**Conclusions**: There is some evidence that being based in the same country as an author may influence a reviewer's decision when the reviewer identity is public. There does however seem to be no evidence that being able to read an existing published review *prior* to submitting your own review overly influences a reviewer's decision or encourages conformity.

These findings are based entirely upon peer review data derived from the F1000Research platform which is one that offers open, invited peer review within a relatively unique post-publication publishing system. The findings do help to dispel some of the critique levelled at aspects of open peer review, though if the data were available we would ideally compare our results with data derived using alternative and established peer review practices (both open and closed). We would like to see a more robust evidence base to support decision making around aspects of the scholarly publishing system and peer review in particular.

**Keywords**: Open science; Peer review; peer review biases; open peer review; F1000Research, post-publication peer review

**Introduction**

The trend towards more open and collaborative knowledge practices ('open science') has led to an increasing number of journals and publishing outlets offering 'open peer review' to various degrees (Lee & Moher, 2017; Morey, Chambers, Etchells, et al., 2016; Ross-Hellauer, 2017).

F1000 today perhaps offers one of the most comprehensive versions of 'open peer review'; across all its publishing platforms, a post-publication peer review model is provided alongside invited, open peer review. Each article, after initial editorial and objective checks (e.g. for ethical approval; plagiarism; and inclusion of underlying data) is published *before* peer review. The subsequent peer review process, notable in the debate about the value of open peer review, publishes the identities and affiliations of all peer reviewers alongside their narrative report. The publication process is iterative ('continuous publishing'), with authors able to respond directly to their reviewers' comments by providing a revised article version, linked to the previous version to preserve the revision history.

To keep the process rapid and iterative, peer reviewer reports are published (following an initial check) as soon as they are submitted thus allowing the possibility for subsequent reviewers to read already published reports before submitting their own review. Additionally, the post-publication peer review model is intended to shift the role of peer review in scholarly publishing from its traditional focus on selecting content for publication, to one that helps to shape the research to be the best it can be – and providing this openly, allows readers to reflect on the development of a piece and consider any reviewer critique. Nevertheless, open peer review introduces the potential for specific types of reviewer biases and since peer review of scholarly output continues to provide an important quality control function (Siler, Lee, & Bero, 2015), it is important to examine the benefits and potential biases that newer open systems of peer review may introduce.

There are many types of bias described within peer review processes and practices operated by grant funding agencies and in scholarly publishing, including according to: author/grant applicant demographic characteristics (Ceci & Williams, 2011; Fox, Burns, & Meyer, 2016; Tamblyn, Girad, Qian & Hanley 2018; Witteman, Hendricks, Straus & Tannenbaum 2019); author-reviewer speciality differences (cognitive distance) (Wang & Sandström, 2015); country of affiliation/article country of origin (Harris, Macinko, Jimenez, Mahfoud, & Anderson, 2015), and perceptions of authors' institutional prestige (Peters & Ceci, 1982). Perhaps because such studies revealing some biases have been typically tested on the single-blind model (with author/grant applicant identity known by reviewers), there remains a widespread belief that double-blind peer review introduces the 'least' bias and is superior to open peer review approaches (Rodríguez-Bravo, Nicholas, Herman, et al., 2017; Mulligan, Hall, & Raphael, 2013; Moylan, Harold, O'Neill, & Kowalczuk, 2014). However, beliefs about the merits of open peer review are supported by limited empirical evidence and based upon survey and attitudinal research (Ross-Hellauer, Deppe, & Schmidt, 2017).

From studies providing empirical insights into open peer review, there is evidence that providing peer review transparency has a minimal effect on review quality. Two randomised control trials of The BMJ did not find quality differences between the reviews provided when reviewers knew that their identities would be revealed, compared to blinded review (Van Rooyen, Godlee, Evans, Black & Smith, 1999) or when reviewers were told their reviews may be published online (Van Rooyen, Delamothe, & Evans, 2010). A study of a psychology journal found that reviews tended to be more carefully written when they were intended to be openly available (Walsh, Rooney, Appleby, & Wilkinson, 2000). A systematic review of research into peer review for biomedical journals found nine studies into concealing author or reviewer identity, but no conclusion could be reached overall about how this influenced review quality (Jefferson, Alderson, Wager, & Davidoff, 2002). However, a recent study of a large volume of articles submitted to Elsevier journals concluded that using an open peer review process does not compromise the quality and participation in the process, at least when reviewers are able to protect their anonymity (Bravo, Grimaldo, López-Iñesta, Mehmani, Squazzoni, 2019), although The BMJ reviewers were less willing to review when their identities would be revealed (Van Rooyen, Godlee, Evans, Black & Smith, 1999). While there is some evidence that open peer review might deter reviewer acceptance rates (Ross-Hellauer, Deppe, & Schmidt, 2017), there is limited empirical evidence about the effect that revealing reviewers' identities and affiliations and 'signed peer review' policies, has on peer review in practice.

As part of the efforts to provide a firmer evidence base to support decision-making around the use of open peer review, this article tests the extent to which a post-publication peer review process, accompanied by fully open and named peer review (used by F1000 across all its publishing platforms) introduces bias. First, since reviewers and authors know each other's identity, is there any evidence that reviewers are more generous to authors who they know or are likely to meet or collaborate with at some point, for example if they are based in the same country. Social factors have been shown to influence peer review decisions in other contexts (regional bias: Walker, Barros, Conejo, Neumann, & Telefont, 2015), and so this is a plausible hypothesis. Second, since reviewers can view and read other reviewer reports before submitting their own for the same article, they may be influenced by what they read (either consciously or subconsciously) and even use the existing report to reduce their own workload. This influence may be greater perhaps if the published reviewer is an established figure within a field. This study investigates evidence for these two types of bias: whether author and/or reviewer country affiliation influences reviewer comments and decisions; and whether the availability of previous reviewer reports influence the comments and decisions of subsequent reviewers.

For academic peer review to be as effective and valuable to scientific discourse as possible, it is essential that potential biases are identified and, as far as possible, their effects mitigated, managed or eliminated. It is also important that scrutiny of newer models is matched with scrutiny of existing and established models.

**Methods**

*Data*

The data consists of the text and judgements of all reviews of version 1 outputs from F1000Research (Kirkham & Moher, 2018) published on the F1000Research platform between July 13, 2012, and July 8, 2019. Reviews were included for all types of articles submitted for peer review. Since the focus is on reviewer bias, there does not seem to be an obvious reason why this should differ between article types. Only reviews posted to first versions of articles were included. This is because reviews after the first version are more likely to give a positive verdict as the authors have revised their article in response to the reviewers' suggestions. Each reviewer report contains mandatory text, and a judgement of "Approved" (no changes or a few small or cosmetic changes), "Approved with reservations" (not fully sound in the current version), or "Not approved" (fundamental flaws and the work overall is poor quality; the authors are still encouraged to revise their article to respond to the concerns raised). Prior to mid-2013, it was possible for a reviewer to submit an overall assessment without any accompanying narrative, however this policy has been revised and now all peer review reports on version 1 must include both a narrative and assessment. All peer review reports and assessments are made public alongside the published article, regardless of whether it is 'approved' or not – the approval status does not affect the decision to publish, but determines when an article is considered to have 'passed peer review' and is therefore sent to indexers such as PubMed and Scopus.

*Author and reviewer affiliation bias tests*

For the author/reviewer affiliation tests, only the country of the first author's affiliation was considered for multi-author papers, since the first author has typically provided the largest range of contributions to the published work. Although authorship can sometimes be misleading (Brand, et al 2015), empirical evidence suggests that the first author tends to be the most substantial contributor in all broad academic fields, even those with senior last author or alphabetical norms (Larivière, Desrochers, Macaluso, Mongeon, Paul-Hus, & Sugimoto, 2016). Thus, for the purposes of this analysis, the first author location was used as a proxy for the location of the whole authorship team.

Only the first peer review report was assessed (for the affiliation tests; all were used for the previous review bias tests) to avoid non-independence for statistical tests due to multiple reports (by different reviewers) on the same article. Multiple reports by different reviewers on the same article are not statistically independent because different judgements of the same article are likely to agree more than different judgements on different articles. Author and reviewer national affiliations were judged from the country of their affiliation, or first affiliation if multiple affiliations were declared. Decisions were coded on the following simple numerical scale to allow averaging.
- 1: Not approved
- 2: Approved-with-reservations
- 3: Approved

Non-parametric Mann-Whitney U tests were used to analyse author/reviewer affiliation bias since the data does not follow the normal distribution (from a Q-Q plot, although its skewness -0.66 and excess kurtosis -1.42 are reasonable).

*Previous review bias tests*

Some reviewer reports specifically made reference to previous reviewer reports, for example to say that they would not repeat points made in them. Such reports were identified by text searches for the strings, "Referee", "referee", Reviewer", and "reviewer", followed by a manual check of whether the term was used to refer to comments posted by a previous reviewer about the same article. This produced 143 examples of reviewers explicitly mentioning a previous reviewer report.

Reviewer reports on a single article were ordered using their identifiers, which are allocated consecutively. The published time of each report was then used to calculate the number of days between reports. A reviewer report is typically published rapidly, often on the same day that it is submitted, although occasionally the editorial team may need clarifications first. If two reports are posted on the same day then it is unlikely but not impossible for the reviewer from the second report to have seen the first report before writing and submitting their own. Nevertheless, the longer the time period between the first report appearing and the second report being published, the more likely it is that a subsequent reviewer had seen a prior report (Barros & Allen, 2017). This is because older reports would have been more likely to be available to view when the subsequent reviewer first accessed the article to read, print, or download.

Agreement rates were calculated for each report after the first. For this, the fraction of previous reports that they agreed with was calculated. For example, if the first and third reviewer assigned an 'Approved' status, and the second reviewer assigned 'Not approved', then the second reviewer would have an agreement rate of 0/1=0 (disagrees with the first reviewer) and the third reviewer would have an agreement rate of 1/2=0.5 (agrees with one of the two prior reviewers).

Statistical tests, as described below, were used to assess the likelihood of bias due to reviewers being influenced by reading prior reviews. Non-parametric tests were used because agreement rates were not normally distributed.

**Results and discussion**

While overall only 5% of articles were assigned a 'Not approved' status during the peer review process, around a third of articles submitted to F1000Research (36% in Jan-Oct 2019 from the authors' knowledge) do not adhere to policies and therefore do not proceed past the initial editor checks and are therefore not sent for peer review; this is similar to estimates of 35%-40% overall reject decisions for academic journals, although 10%-15% lower for open access and health/biomedical journals (Björk, 2019; Sugimoto, Larivière, Ni, & Cronin, 2013).

*Do author and reviewer national affiliations influence peer review judgements?*

The first area of potential bias is whether peer reviewers might be inclined to provide a favourable review to authors based in the same country, especially when their identity will be revealed. As background for this, the average judgements were analysed based on the national affiliations of authors and reviewers separately.

**First, do authors from different countries receive different average judgements?** Average judgements for articles varied with first author affiliation country to some extent (Figure 1). For example, decisions are statistically significantly more positive for the UK and USA than for India. This aligns with previous research showing that there are international differences in peer review outcome for authors from different regions (Walker *et al*, 2015) and in the average citation impact of academic research (Smith, Weinberger, Bruna, &

Allesina, 2014), although it is influenced by collaborations with higher income and research-intensive countries (Thelwall & Levitt, 2018). The main reasons cited for this potential inequality in research intensity and citation impact between countries include differing levels of financial support for research infrastructure, limited availability of mentors and experienced researchers, and the migration of highly skilled researchers to better-resourced economies (Xie 2014).

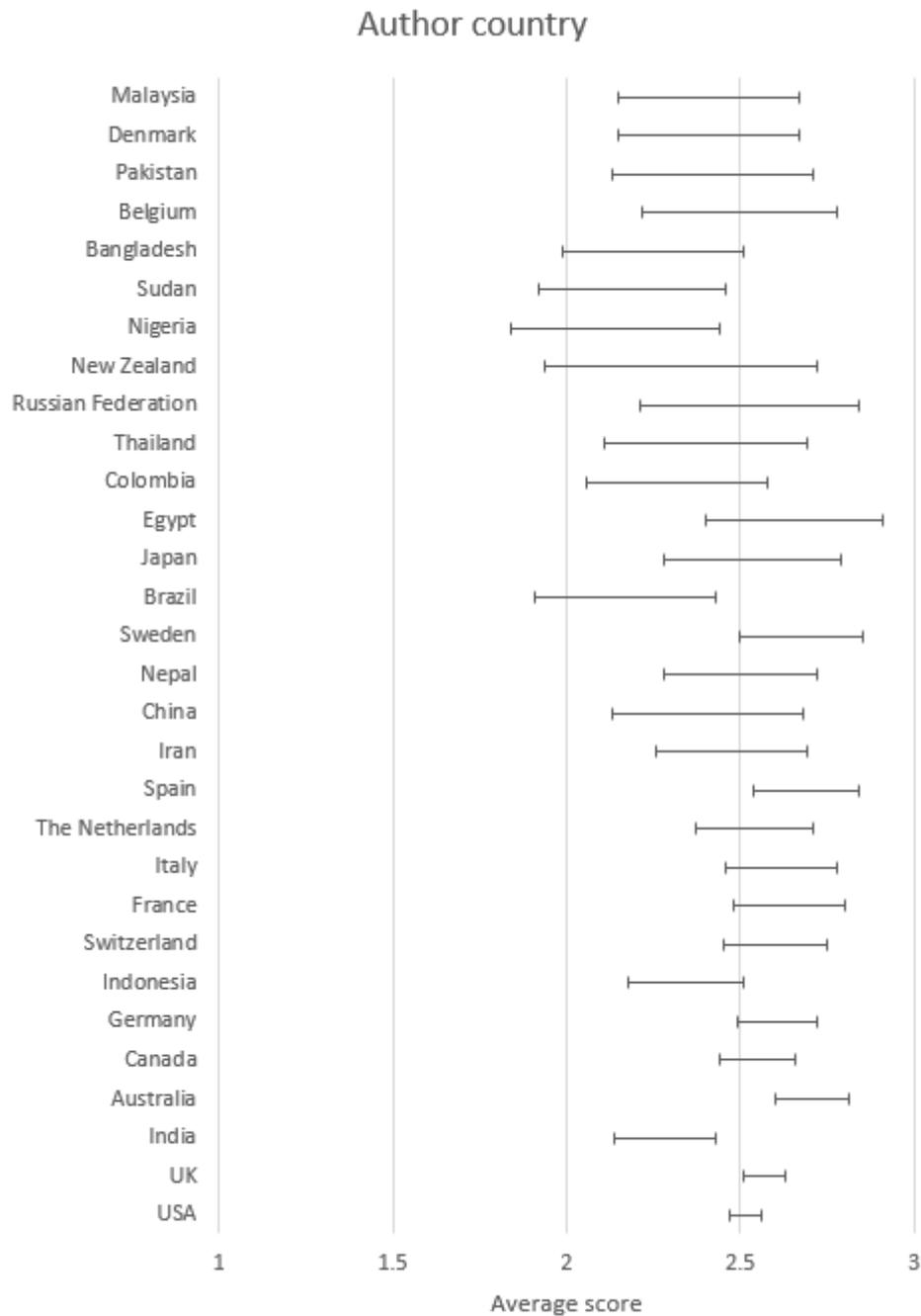

Figure 1. 95% confidence intervals for the average judgements on articles by the first reviewer, split by the first affiliation country (top 30 countries) of the first **author**. Countries are in order of number of articles.

**Second, do reviewers from different countries deliver different average judgements?** Reviewer judgements largely did not vary according to the location of the reviewer, with the variations found being of a level to be expected by normal statistical variations since the confidence intervals mainly overlap and all except two contain the 2.5 line (Figure 2). Thus, there do not seem to be international variations in the leniency of reviewers in general. We noted two possible exceptions in our data set: reviewer reports from Egypt and Turkey seem to be more positive than normal.

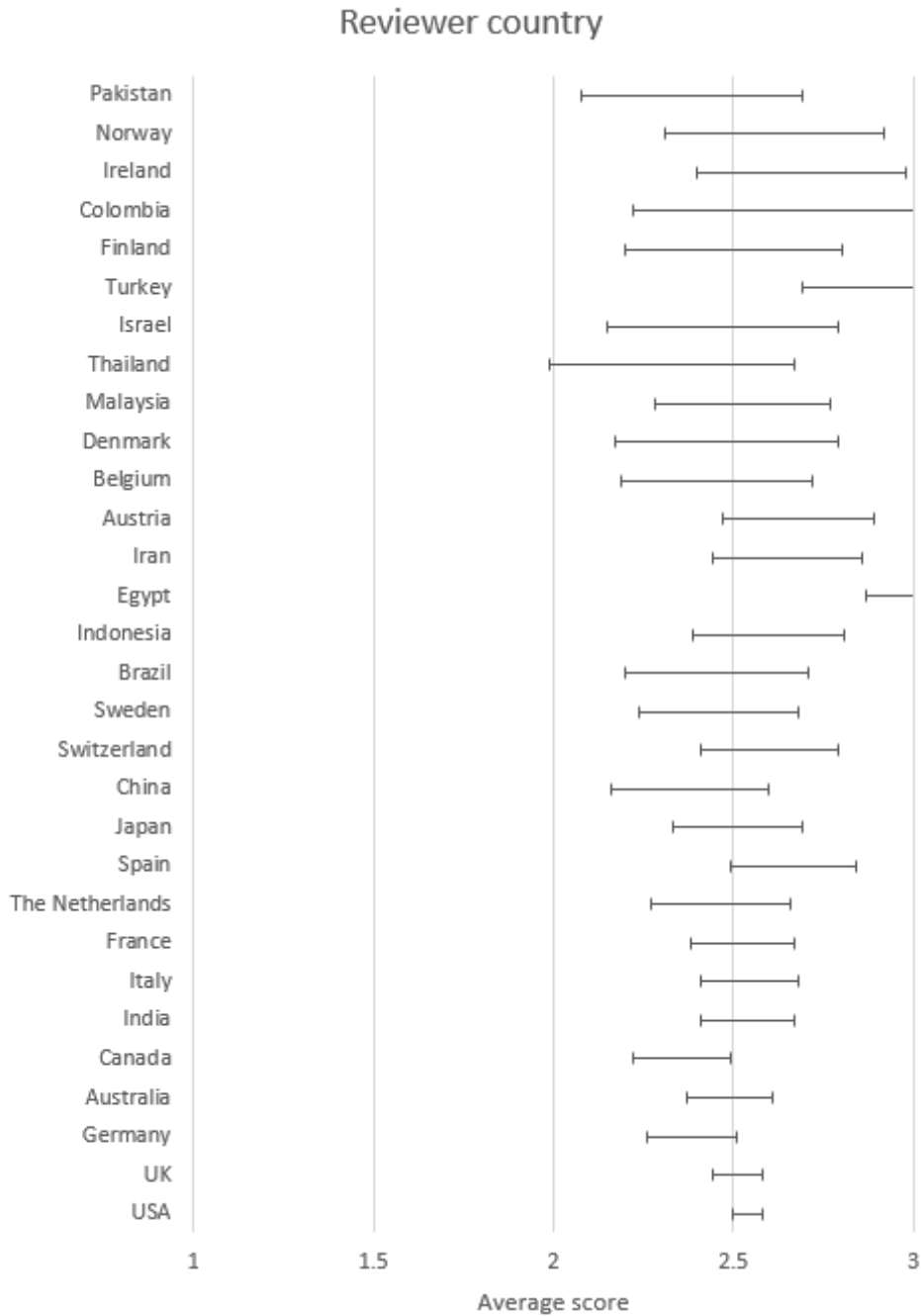

Figure 2. 95% confidence intervals for the average judgements on articles by first **reviewer** affiliation country (top 30 countries). Countries are in order of number of reviewers.

**Third, do countries with authors that receive more positive judgements also tend to give more positive judgements?** There is a slight tendency for countries that allocate more positive judgements to also receive more positive judgements (Pearson correlation 0.13 for Figure 3; Pearson correlation 0.15 for the top 30). Although this is clearly true for Spain in contrast to Brazil, the low correlation suggests that this is not a strong international trend.

The weak correlation could be due to international differences in fields submitted and reviewed for F1000Research, since there may be disciplinary differences in reviewing strictness. The positions of Spain and Brazil could be due to differing topics submitted and reviewed, for example. Similarly, it could also be a difference in the breakdown of article types received from different countries, as some article types – for example Research Articles – require more stringent reviewing than others.

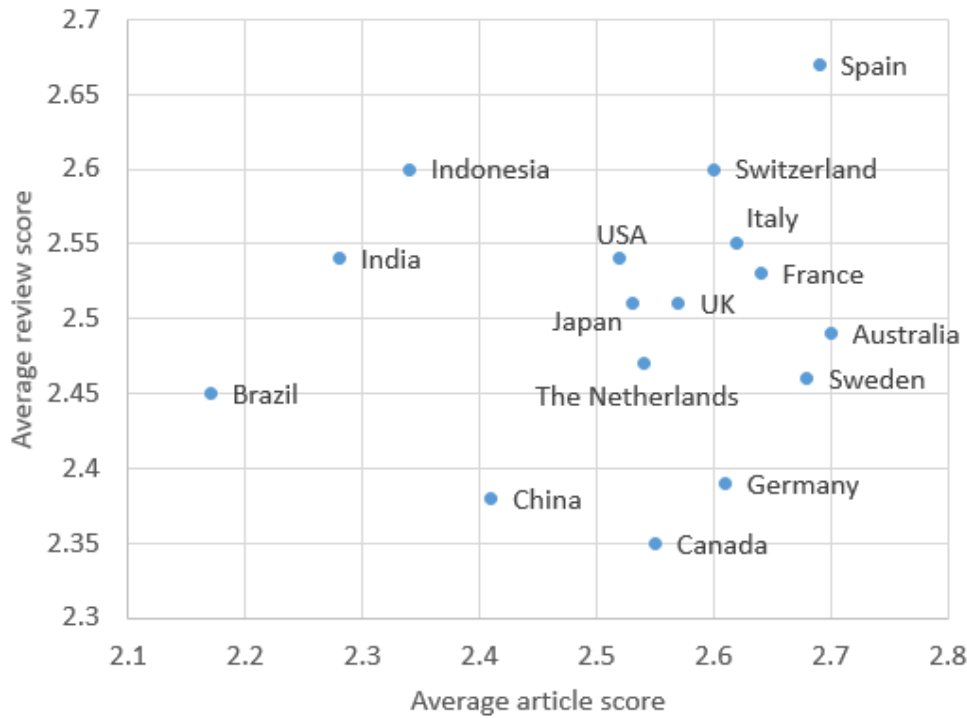

Figure 3. Average judgements on articles by first reviewer affiliation country against first affiliation country of the first author (the 17 countries with at least 30 articles and at least 30 reviewers).

**Fourth, do authors receive more positive judgements from reviewers with the same national affiliation?** This is the main test for this section. For each country, Mann-Whitney U tests were run to see whether articles tended to receive more positive judgements if the reviewer was affiliated with an institution in the same country as the first author (Table 1). A degree of *same* country bias existed for 16 of the 20 countries with the most articles (one country, Belgium, could not be tested due to a lack of same-country reviewers). After a Sidak correction, the difference was statistically significant in one case, Egypt, and close to significant in one other, the UK. The slightly more powerful Benjamini-Hochberg correction method (Benjamini & Hochberg, 1995) would not change these results.

Reviewers were substantially more positive about Egypt-based articles if they are also based in Egypt (a score of 3: all 15 first reviewer reports from Egypt-based reviewers assigned an 'Approved' status on an Egypt-based first-authored article) compared to if the reviewer was based outside Egypt (2.29 from 23 reports). For second, third, fourth, and fifth reviewer reports, there were 14 by Egypt-based reviewers, 13 of whom assigned an 'Approved' status and one 'Approved-with-reservations'. Overall, the results suggest a community of supportive Egyptian dentistry-related researchers actively publishing in F1000Research (see Box 1). In contrast, the UK results (see Box) might be due to the country submitting strong medicine-related articles (see Box 2).

**Box 1: Investigations of apparent Egyptian reviewer bias**

Only one out of 29 reports on Egypt-based author articles was assigned anything other than an 'Approved' status by an Egypt-based reviewer. Most of the author and reviewer affiliations were associated with dentistry. A possible explanation is that Egypt has excellent dentistry research, submits its high-quality dentistry research to F1000Research or has a quality control procedure in dentistry to ensure that articles are fully checked. This seems unlikely because other reviewers are less positive. For one research article, the obviously knowledgeable Egyptian reviewer recommends Approved, but the also knowledgeable UK reviewer makes a good case for Not approved (https://f1000research.com/articles/6-1786/v1). For another research article, the Egyptian reviewer report does not seem strong and the US reviewer makes a convincing case that the article should not be approved (https://f1000research.com/articles/7-1703/v1). Thus, there may be a supportive community of Egyptian dentistry researchers that express their support by not being overly critical of compatriots or by recommending less experienced colleagues as reviewers.

**Box 2: Investigations of possible UK reviewer bias**

Here is an example where the UK reviewer assigned an 'Approved' status and the two non-UK-based reviewers assigned 'Approve-with-reservations'. The corrections suggested by all three reviewers seem minor, so all decisions could have been Approved (https://f1000research.com/articles/6-170/v2). Another example has two UK Approveds and one non-UK Approved-with-reservations from a more detailed review, which seems reasonable (https://f1000research.com/articles/7-55/v2). Another example has one UK Approved and two non-UK Approved-with-reservations, all of which seem reasonable decisions (https://f1000research.com/articles/7-1133/v2). In all cases where a non-UK reviewer assigns 'Not approved' for an article, at least one UK reviewer also assigns 'Not approved', so there are no obvious cases of bias. Thus, there is possibly a slight tendency for UK reviewers to err on the side of generosity on the borderline of Approved and Approved-with-reservations for articles first authored by compatriots. There were also many examples of UK articles with UK-only based reviewers, all of which assigned an 'Approved' status (e.g., https://f1000research.com/articles/7-714/v1 https://f1000research.com/articles/7-1107/v1), so it is *possible* that there is a medicine-related tendency for UK based authors to submit strong articles to F1000Research and suggest UK researchers as reviewers.

Table 1: Mann-Whitney U tests for same-country reviewer bias (the 20 countries with at least 20 articles). A Sidak correction gives a critical value of p=0.0026 to preserve the familywise error rate at 0.05. There were no same-country reviewers for Belgium.

| First author country | Articles | Mean decision score | | Difference | P* |
| --- | --- | --- | --- | --- | --- |
| | | Same country reviewer | Different country reviewer | | |
| USA | 737 | 2.53 | 2.50 | 0.03 | 0.6197 |
| UK | 312 | 2.68 | 2.51 | 0.18 | **0.0077** |
| India | 110 | 2.63 | 2.19 | 0.44 | **0.0147** |
| Australia | 105 | 2.93 | 2.67 | 0.27 | 0.0684 |
| Canada | 104 | 2.38 | 2.57 | -0.19 | 0.2976 |
| Germany | 99 | 2.53 | 2.62 | -0.09 | 0.6448 |
| Indonesia | 70 | 2.61 | 2.21 | 0.40 | **0.0387** |
| Switzerland | 57 | 3.00 | 2.58 | 0.42 | 0.2918 |
| Italy | 55 | 2.72 | 2.57 | 0.15 | 0.4917 |
| France | 55 | 2.71 | 2.63 | 0.09 | 0.8269 |
| The Netherlands | 52 | 3.00 | 2.50 | 0.50 | 0.0991 |
| Spain | 39 | 3.00 | 2.64 | 0.36 | 0.0840 |
| Iran | 38 | 2.67 | 2.30 | 0.37 | 0.1162 |
| China | 32 | 2.83 | 2.31 | 0.53 | 0.1299 |
| Sweden | 31 | 2.40 | 2.73 | -0.33 | 0.1638 |
| Japan | 30 | 2.75 | 2.50 | 0.25 | 0.5670 |
| Brazil | 30 | 2.00 | 2.18 | -0.18 | 0.6821 |
| Egypt | 29 | 3.00 | 2.29 | 0.71 | **0.0024** |
| Thailand | 25 | 2.40 | 2.40 | 0.00 | 0.9755 |
| Belgium | 20 | None | 2.5 | NA | NA |

*Bold values are statistically significant without a familywise error rate correction; bold underlined values are also statistically significant with a familywise error rate correction. Red values indicate a negative difference in decision scores.

### *Does visibility and access to another reviewer's report influence subsequent reviewers?*

The second area of potential bias explored is whether through an open, non-anonymised peer review process, reviewers' decisions are influenced by the ability to view the report of another reviewer.

*First, are reviewer reports more likely to refer to previous reviewer reports if the time period between the first report appearing and the second report being published is longer?* This question needs a positive answer to justify the main test assumptions. This was assessed with a Mann-Witney U test for a difference in average time (rank) between reports for two groups:
- Group 1: The subsequent reviewer report did not mention any prior reviewer report.
- Group 2: The subsequent reviewer report mentioned a prior reviewer report.

The Mann-Whitney U test found very strong evidence (p=0.001) to reject the null hypothesis that there was no difference between the groups. Group 2 had higher ranks (longer time periods). Thus, there is very strong evidence that longer periods between reviewers associate with the subsequent reviewer being more likely to refer to a previous review. This does not prove that subsequent reviewers are more likely to have read a previous report if it was published longer ago because many reviewers may have read previous reports and not mentioned them. Nevertheless, this is a plausible hypothesis, both from common sense (as argued in the Methods section) and from this secondary test.

***Second, are reviewers more likely to mention a previous report for any outcome?*** A reviewer might only mention a previous report to discuss its criticisms, for example. A chi-square test was used to test for this, finding very strong evidence (p<0.001) to reject the null hypothesis that there is no relationship between the decision and whether a previous report is mentioned. The main reason for this is that a reviewer report that had assigned an 'Approved-with-reservations' status was more likely to mention a previous reviewer report than was a report that had assigned an 'Approved' status (Table 2). This is likely to be because there are more potential issues for discussion and for authors to consider, or to avoid reiterating previously discussed issues.

Table 2. Chi square test for a relationship between a specific reviewer assignment and whether a previous reviewer report is mentioned. Reports on version 1 publications in F1000Research (all article types).

|  |  | Mentions a previous report | | Total |
|---|---|---|---|---|
|  |  | No | Yes |  |
| Not approved | Reports | 164 | 5 | 169 |
|  | Expected Reports | 161.5 | 7.5 | 169 |
| Approved with reservations | Reports | 1075 | 82 | 1157 |
|  | Expected Reports | 1105.9 | 51.1 | 1157 |
| Approved | Reports | 1855 | 56 | 1911 |
|  | Expected Reports | 1826.6 | 84.4 | 1911 |
| Total | Reports | 3094 | 143 | 3237 |

***Third, are there variations in the time taken for a reviewer to assign a given article approval status?*** A Kruskal-Wallis test was used to compare the time taken to provide a peer review for the three approval statuses allowed in the F1000Research peer review model. There was a significant difference (p=0.001) between decision times, with more positive approval statuses being assigned in a shorter time (Table 3). This is probably due to the challenge of securing sufficient peer reviewers to review more complex or controversial work that needs substantial reviewing effort. The time taken between an article being published and a first review report appearing and it being assigned a 'Not approved' status took on average twice as long as a for an article where the first report was assigned an 'Approved' status.

Table 3. Reviewer status assignments and the time between an article and its reviews (n=5821) being published (calculated separately for each review). Reports on version 1 publications in F1000Research of all article types.

| Decision | Reviews | Average (geometric mean) days to review |
|---|---|---|
| Not approved | 341 | 48.6 |
| Approved with reservations | 2052 | 36.8 |
| Approved | 3428 | 24.2 |

***Fourth, are reviewers more likely to agree with a prior reviewer report if it is mentioned in their review?*** A Mann-Whitney U test was used to compare the agreement rates between reviewers who did and did not mention previous reviewers. The mean ranks were almost the same, so there is insufficient evidence to reject the null hypothesis that there is no difference (p=0.956). Based on the similar mean ranks, it seems unlikely that there is a substantial influence of prior reviewers on subsequent reviewers that mention their reports.

***Fifth, are reviewers more likely to agree with prior reviewer reports when they have had longer to read them?*** This is the main test for this section. The above test is not conclusive because reviewers may avoid mentioning previous reports that they have read but disagree with. Since it is not known whether a reviewer has read a previous report, an indirect test was run, using the time between reports as an indicator of the likelihood of a reviewer having read a previous report, as argued for and tested for above (first point). A Spearman correlation test was used to compare reviewer agreement rates with the time since the previous review report. There was a statistically significant negative correlation (-0.123, p<0.001), giving strong evidence to reject the null hypothesis that there is no relationship. There is thus a negative relationship: longer gaps (more time to read a review) associated with a slightly higher rate of *disagreement*. This is not conclusive evidence of the absence of a conformity bias from reviewers reading previous reports because the more controversial papers may require extra reviewers or may need longer for reviewing or recruitment. Nevertheless, in conjunction with the neutral result for reports mentioning previous reviewer reports, it suggests that the availability of previous reports does not produce a strong conformity pressure on many reviewers.

**Conclusions**

While the use of peers as experts to provide quality assurance for scholarly output remains a cornerstone of modern science, it is known to introduce bias and is ultimately subjective. One of the arguments for the shift to more open models of peer review is that by providing comments and reviewer identities in the open, readers can consider experts' perspectives of the strengths and weaknesses of an article. Furthermore, it is argued that requiring reviewers to provide comments in the open will help to ensure that peer review is constructive, requiring reviewers to justify negative judgements, potentially reducing the scope for conscious bias.

This study found a tendency for reviewers to be more likely to assign a positive review to authors based in the same country as them. The mostly likely explanation is bias due to reviewers trying to help (or avoid animosity with) people that they are more likely to know or meet. Nevertheless, reviewers might also decline to review rather than publishing a public negative review for an acquaintance. In support of this, a study of psychology found that reviewers were more willing to sign positive reports (Walsh, Rooney, Appleby, & Wilkinson, 2000), so reviewers may have concerns about authors knowing their identity for negative reviews. It is not known whether a location bias is common practice among authors and reviewers more generally, due to the absence of comparable data among publishers using traditional, closed peer review. Moreover, there may be similar correlations between grant applicants and their reviewers according to location (though the reverse has been found for Australian grants: Marsh, Jayasinghe, & Bond, 2008). Thus, whilst this study has found evidence of an affiliation bias, it is not clear whether the bias would also occur in single blind review (i.e., whether it is due to the author's affiliation being public or the reviewer's affiliation being public).

In terms of whether the ability to view the report of one reviewer exerts an influence on subsequent reviewers, there was little evidence that this is the case. It seems unlikely that the prior public availability of reviewer reports creates a pressure on reviewers towards conformity. The slight tendency towards non-conformity found in one test may well be a side-effect of review complexity rather than a genuine non-conformity bias.

As noted, it is important that the introduction of new approaches to peer review are tested to ensure that they do not have unintended consequences nor introduce new biases. Nevertheless, scrutiny of new processes needs to be accompanied by fuller scrutiny of established processes; we have been unable to compare our findings with the extent of bias along the dimensions we explored within other peer review models (e.g. single-blind) due to a lack of data. To be able to refine and optimise the use of experts (e.g. peer review) in grant funding and scholarly publishing requires closed peer-review data to be made available in formats that enable comparable analysis. We fully acknowledge that the findings of this study are tentative at best, weakened by our reliance upon hypotheses that cannot be fully tested and an inability to control for all relevant independent variables, such as specialist areas. A randomised controlled trial would be needed to fully assess the influence of open reviewer identities. Without this, the results may be due to unmodelled variables, such as international differences in the specialties of articles (for national affiliation bias) and the possibility that reviewers tend to ignore published reports that they disagree with (for prior report publication bias).

We hope that this analysis will contribute to the evidence base and decision-making about how and where open peer review can be used to best effect. We particularly hope to see a growth in studies that aim to understand how the processes and workflows used to support research can influence the outcomes of research itself.